\documentclass[conference,letterpaper,10pt,twocolumn]{ieeeconf}
\IEEEoverridecommandlockouts                              
\bstctlcite{IEEEexample:BSTcontrol}

\usepackage{scalefnt}
\usepackage{setspace}
\usepackage{romannum}
\usepackage[usenames,dvipsnames]{xcolor}
\usepackage{tkz-berge}
\usetikzlibrary{fit,shapes}
\usepackage{calrsfs}
\usepackage{graphicx}
\usepackage{graphics}
\usepackage{epsfig}
\usepackage{mathptmx}
\usepackage{caption}
\usepackage{subcaption}
\usepackage{times}
\usepackage{tikz}
\usetikzlibrary{positioning}
\usepackage{amsmath, amssymb}
\usetikzlibrary{arrows}
 \usepackage[T1]{fontenc}
\usepackage{bm, bbm}

\usepackage{kantlipsum}
\usepackage{amsthm}
\usepackage{amsfonts}
\usepackage{setspace}
\usepackage{multirow}
\usepackage{textcomp}
\usepackage[english]{babel}
\usepackage{float}
\usepackage{algorithmicx}
\usepackage[section]{algorithm}
\usepackage{algpascal}
\usepackage{algc}
\usepackage{algcompatible}
\usepackage{multicol}
\usepackage{algpseudocode}
\let\oldReturn\Return
\renewcommand{\Return}{\State\oldReturn}
\usepackage{mathtools}
\usepackage{bm}
\usepackage{mathptmx}
\usepackage{times}
\usepackage{mathtools, cuted}
\usepackage{textcomp}
\usepackage[english]{babel}
\usepackage{rotating}
\usepackage{pgfplots}
\pgfplotsset{compat=1.5}
\usepackage{siunitx}
\usepackage{pgfplotstable}
\usepackage{filecontents}

\usepackage{bbm}
\usepackage{bbold}
\usepackage{float}
\def\BibTeX{{\rm B\kern-.05em{\sc i\kern-.025em b}\kern-.08em
    T\kern-.1667em\lower.7ex\hbox{E}\kern-.125emX}}

\newtheorem{prop}{Proposition}

\newtheorem{assume}{Assumption}

\DeclareMathAlphabet{\pazocal}{OMS}{zplm}{m}{n}

\newcommand{\norm}[1]{\left\lVert#1\right\rVert}
\newcommand{\G}{\pazocal{G}}
\newcommand{\N}{\pazocal{N}}

\newcommand{\bX}{{\bf X}}
\newcommand{\bXs}{{\bf X}^{\star}}
\newcommand{\Y}{{\pazocal{ Y}}}
\newcommand{\bU}{{\bf U}}
\newcommand{\bUs}{{\bf U}^{\star}}
\newcommand{\bV}{{\bf V}}
\newcommand{\bA}{{\bf A}}
\newcommand{\bB}{{\bf B}}
\newcommand{\bBs}{{\bf B}^{\star}}
\newcommand{\bZ}{{\bf Z}}

\newcommand{\ba}{{\bf a}}
\newcommand{\bb}{{\bf b}}

\newcommand{\bx}{{\bf x}}
\newcommand{\bxs}{{\bf x}^{\star}}
\newcommand{\by}{{\bf y}}

\newcommand{\be}{{\bf e}}

\newcommand{\A}{\mathbf{A}}

\newcommand{\qreq}{\overset{\mathrm{QR}}=} 
\newcommand{\svdeq}{\overset{\mathrm{SVD}}=} 

\def \*{\star}

                     \makeatletter
\makeatletter
\newcommand{\specificthanks}[1]{\@fnsymbol{#1}}
\makeatother
\title{\LARGE \bf Fully Decentralized and Federated  Low Rank  Compressive Sensing }
\author{Shana Moothedath and Namrata Vaswani\thanks{\textsuperscript{$^*$}Department of Electrical and Computer Engineering, Iowa State University, USA. Email: $\lbrace$mshana,namrata$\rbrace$@iastate.edu.}
}

\begin{document}
\maketitle

\begin{abstract}
In this work we develop a fully decentralized, federated, and fast solution to the recently studied Low Rank Compressive Sensing (LRCS) problem: recover an $n \times q$ low-rank matrix $\bXs = [\bxs_1, \bxs_2, \dots, \bxs_q]$ from column-wise linear projections, $\by_k : = \bA_k \bxs_k, \ k=1,2, \dots, q$, when each $\by_k$ is an $m$-length vector with $m < n$. An important application where this problem occurs and a decentralized solution is desirable is in federated sketching: efficiently compressing the vast amounts of distributed images/videos generated by smartphones and various other devices while respecting the users' privacy. Images from different devices, once grouped by category, are pretty similar and hence the matrix formed by the vectorized images of a certain category is well-modeled as being low rank.  A simple federated sketching solution is to left multiply the $k$-th vectorized image by a random matrix $\bA_k$ and to store only $\by_k$. When $m \ll n$, this requires much lesser storage than storing the full image, and is much faster to implement than traditional image compression. Suppose there are $p$ nodes (say $p$ smartphones), and each stores a set of $(q/p)$ sketches of its images.
We develop a decentralized projected gradient descent (GD) based approach to jointly reconstruct the images of all the phones/users from their respective stored sketches. The algorithm is such that the phones/users never share their raw data (their subset of $\by_k$s) but only summaries of this data with the other phones at each algorithm iteration. Also, the reconstructed images of user $g$ are obtained only locally. Other users cannot reconstruct them. Only the column span of the matrix $\bXs$ is reconstructed globally.
By ``decentralized'' we mean that there is no central node to which all nodes are connected and thus the only way to aggregate the summaries from  the various nodes is by use of an iterative consensus algorithm that eventually provides an estimate of the aggregate at each node, as long as the network is strongly connected.
We validated the effectiveness of our algorithm via extensive simulation experiments.
\end{abstract}
\section{Introduction}\label{intro}

Due to the growing need for reliable high-speed computing and  the increasing focus on security and privacy, it is often preferred to store and process  data in a distributed manner, and to recover the whole data in a federated way \cite{srinivasa2019decentralized, hughes_icml_2014}.
{\em Federated learning} is an approach where devices collaborate to learn a global model from data stored on distributed  devices, under the constraint that device-generated data are stored and processed locally, with only intermediate updates being shared between the devices \cite{kairouz2019advances, bonawitz2019towards}. In the traditional federated setting, so-called as {\em centralized federated learning}, the devices periodically communicate their  local  intermediate updates  with a central server.  The central server  then aggregates  the information received from all the devices and communicates it with all devices. The key limitation of a centralized federated  setting  is that (i)~the central server orchestrates the whole process and hence is a single point of failure and (ii)~the central server may become a bottle neck in certain applications as the number of nodes increases.
In applications such as  federated sketching of data/images from smart phones or IoT devices, a decentralized setting is more practical \cite{savazzi2021opportunities}. This motivates a fully decentralized and federated  framework to learn from distributed data. By ``decentralized'' we mean that there is no central node to which all nodes are connected and thus the only way to aggregate the summaries from  the various nodes is via information exchange between the nodes.

In this paper, we develop a fully decentralized solution to the recently studied Low Rank Compressive Sensing (LRCS) problem \cite{lrpr_gdmin}: how to reconstruct a low rank (LR) matrix from linear projection measurements of its columns in a decentralized  and federated setting.
Specifically, an $n \times q$ (low rank) rank-$r$ matrix, $ \bXs = [\bxs_1, \bxs_2,\ldots, \bxs_q]$, needs to be recovered from distributed column-wise linear measurements $\by_1, \by_2, \ldots, \by_q$, where $\by_k := \bA_k \bxs_k$ for $k \in \{1,2,\ldots, q\}$, and each $\by_k$ is an $m$-length vector with $m < n$. 
The measurement signals, $\by_1, \by_2, \ldots, \by_q$, are distributed across $p$ nodes and the nodes  collaborate to recover $\bXs$ by periodically sharing their local information with  the neighboring nodes via a communication network.  Moreover, the information sharing is federated  such that the nodes share the  parameters of their local model, rather than the raw signal itself.
An important application where this problem occurs and a decentralized solution is needed is for the federated sketching \cite{hughes_icml_2014, li2020federated}, \cite{ wainwright_linear_columnwise, cov_sketch} problem described in the abstract. 

\subsection{Related Work}
The centralized LRCS problem has  been studied in three recent works. The first is an Alternating Minimization solution that solves the harder magnitude-only generalization of LRCS (LR Phase Retrieval) \cite{lrpr_tsp,lrpr_it,lrpr_best}. The second (parallel work) studies a convex relaxation called mixed norm min \cite{srinivasa2019decentralized}. The third \cite{lrpr_gdmin} is a gradient descent (GD) based provable solution to LRCS, that we called GDmin. The convex solution is very slow, has very bad experimental performance, and has a worse sample complexity than GDmin for highly accurate recovery settings \cite{lrpr_gdmin}. The AltMin solution \cite{lrpr_tsp,lrpr_it,lrpr_best}. is also  much slower than GDmin. Also, since it is designed for a harder problem, its sample complexity guarantee for LRCS is sub-optimal compared to that of GDmin, and consequently it has worse recovery performance with fewer samples \cite{lrpr_gdmin}. While \cite{lrpr_gdmin} considered federated, it is the centralized setting where a central node  aggregates  information from all nodes (Figure~\ref{fig:central}).

We should mention here that LRCS is significantly different from the other more commonly studied LR recovery problems: LR matrix sensing (LRMS) \cite{lowrank_altmin}, LR matrix completion (LRMC) \cite{lowrank_altmin, matcomp_candes},  multivariate regression (MVR) \cite{wainwright_linear_columnwise}, or robust PCA \cite{rpca2}. MVR is the LRCS problem with $\A_k = \A_1$, for $k \in \{1,2,\ldots, q\}$,  but this simple change makes it a very different problem: with this change, the measurements of the different columns are no longer mutually independent, conditioned on $\bXs$. This, in turn, implies that the required sample complexity per column,  $m$, can never be less than the signal length $n$. This point is explained in detail in \cite{lrpr_gdmin}.

Distributed iterative algorithms for  consensus and averaging problems are well studied in the literature \cite{xiao2007distributed, olfati2004consensus, olshevsky2009convergence} and they find applications in wide range of areas including distributed agreement and
synchronization problems \cite{lynch1996distributed},  load balancing \cite{cybenko1989dynamic}, distributed coordination of mobile autonomous agents \cite{jadbabaie2003coordination}, and distributed data fusion in sensor networks \cite{xiao2005scheme, spanos2005distributed}. It has been shown in \cite{olfati2004consensus} that a directed graph solves the average-consensus problem using a linear protocol if and
only if it is balanced, i.e., in-degree equals out-degree for all nodes in the   network.
Later \cite{dominguez2011distributed} proposed distributed strategies for average consensus in a directed network. Recently, consensus-based approaches were proposed for decentralized deep learning in \cite{kong2021consensus}, where the notion of consensus distance was used to analyze the gap between centralized and decentralized deep learning models. Reference \cite{savazzi2021opportunities} proposed a consensus-based distributed Machine Learning  (ML) approach for automated industrial systems, where the agents exchange   local ML parameters and update them sequentiallly to learn a global ML model. In this paper, we utilize consensus algorithm to exchange parameters of local projected gradient  between neighboring agents in order to learn or recover an unknown data matrix.

\subsection{Our Contribution and Paper Organization}
In this work, we develop a fast Gradient Decent (GD) algorithm for solving the decentralized federated LRCS problem. The centralized federated setting considered in earlier work \cite{lrpr_gdmin} meant that the algorithm for dealing with the distributed nodes was pretty straightforward and not too different from what would be done in a fully centralized setting. For example, there was no change to the analysis and almost no extra steps needed in the algorithm itself when compared with a fully centralized setting.
However, without any central server to aggregate the summary statistics, the algorithm design becomes much more difficult.  In this work, we borrow ideas from the scalar consensus literature \cite{xiao2007distributed} to develop (i)~a decentralized spectral initialization approach; and (ii)~develop a decentralized approach to aggregate the gradients. Our proposed algorithm, DeF-GD,  integrates a consensus-based approach with projected GD.  We present  numerical validation of our approach through extensive experiments on simulated data.

The rest of the paper is organized as follows.  Section~\ref{sec:prob} introduces the  problem formulation and discusses the notations used in the paper. 
Section~\ref{sec:algos} presents the proposed algorithm, DeF-GD. Section~\ref{sec:sim} gives the numerical validation results of the proposed algorithm. Section~\ref{sec:con},  presents the concluding remarks and future work.

\section{Problem Formulation and Notations}\label{sec:prob}
\subsection{Problem Formulation: DLRCS Problem}
We first specify the  LRCS problem below and then explain the decentralized setting. The goal is to recover a set of $q$ $n-$dimensional vectors/signals, $\bxs_1, \bxs_2, \ldots, \bxs_q$ such that the $n \times q$ matrix  ${\bf X}^{\star}:= [\bxs_1, \bxs_2, \ldots, \bxs_q]$ has rank $r \ll \min(n,q)$,  from column-wise linear measurements of the form
\begin{equation}\label{eq:PR-1}
\by_{k}:= \bA_k ~\bxs_k, ~k=1,2,\ldots, q.
\end{equation}
Here the matrices $\bA_k \in \mathbb{R}_{m \times n}$ are known, $\mathbb{R}$ denotes the set of real numbers, and  $\by_k$ is an $m$-length vector.  We refer to  ${\bf X}^{\star}$ as a Low Rank (LR) matrix as  $r \ll \min(n,q)$.

In this work, we assume a decentralized federated setting. The $q$ signals $\by_1, \by_2, \ldots, \by_q$ are not sensed/measured  centrally at one node. Instead, there is a  set of $p$ distributed nodes/sensors and each node can  observe $q/p$ linear projection measurements. For simplicity we assume here that $q/p$ is an integer.  Thus, for example, node~1 observes $\by_1, \ldots, \by_{(q/p)}$, node~2 observes $\by_{(q/p) +1}, \ldots, \by_{2q/p}$, and so on.

\begin{figure}[h]
\centering
 \begin{subfigure}{.49\linewidth}
 \centering
	\includegraphics[width=\textwidth]{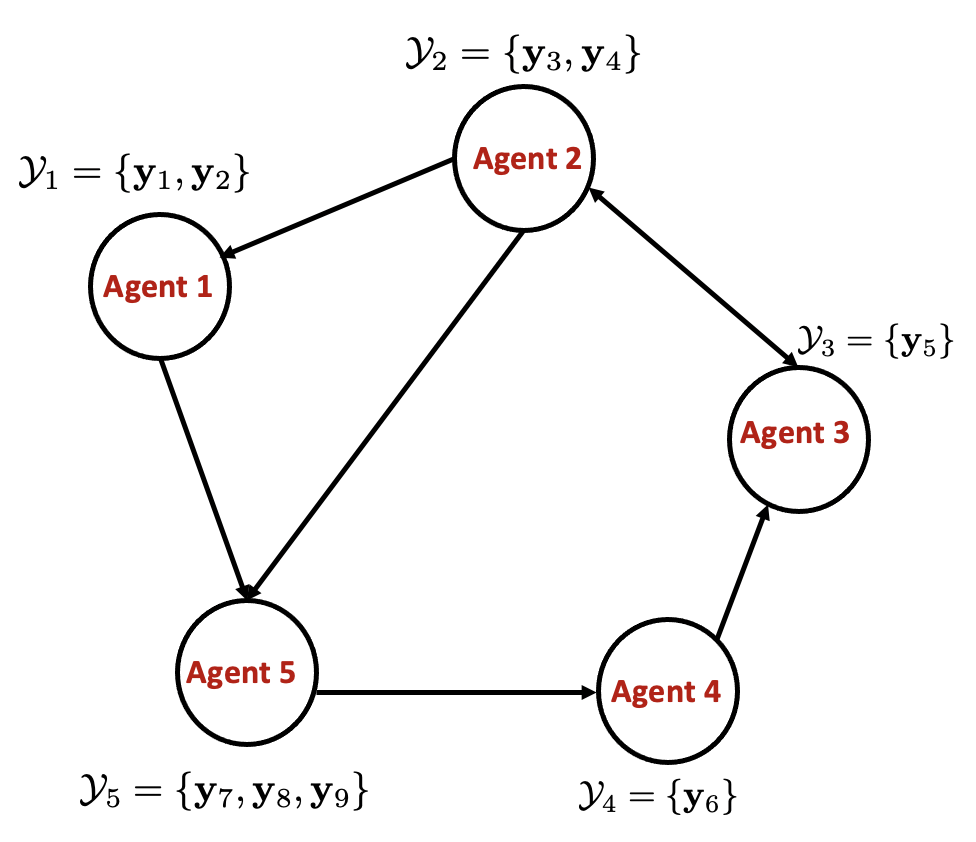}
	\caption{}
	\label{fig:decentral}
	\end{subfigure}
 \begin{subfigure}{.49\linewidth}
  \centering
	\includegraphics[width=\textwidth]{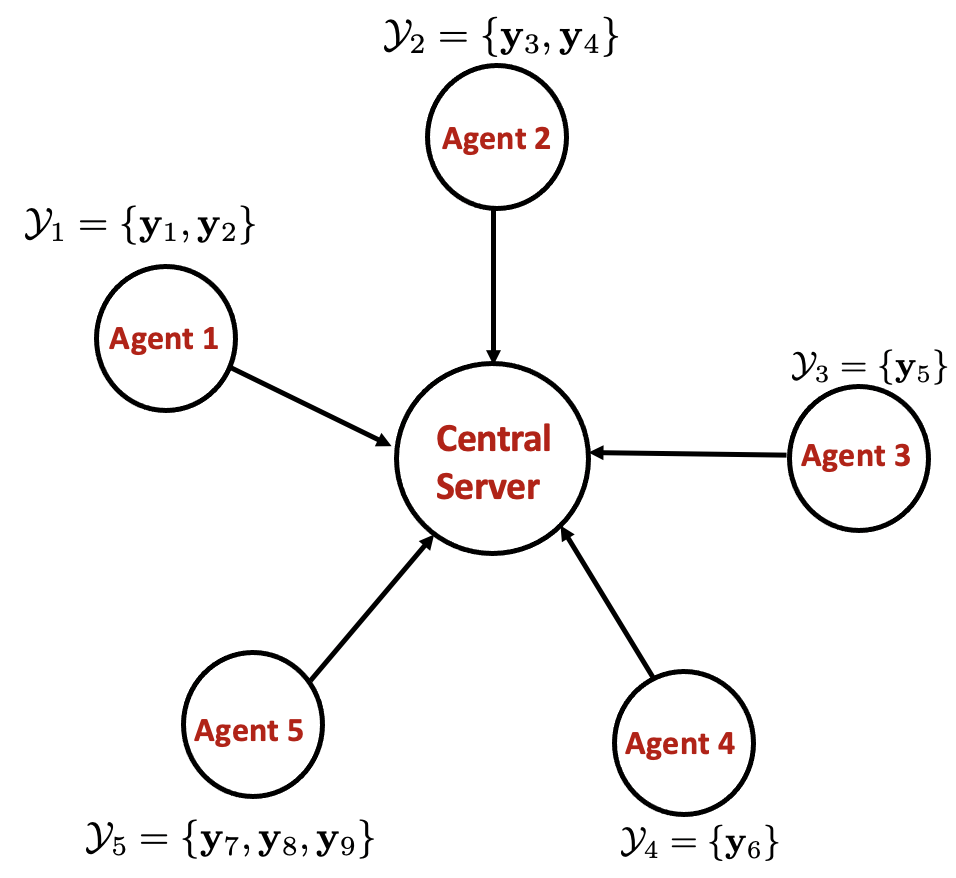}
	\caption{}
	\label{fig:central}
	\end{subfigure}	
\caption{Schematic diagram of the data distribution architecture; (a)~decentralized architecture considered in this work where nodes only share their local update information with their neighboring nodes and (b)~distributed architecture where all nodes share the information of their local updates with a central server considered in \cite{lrpr_gdmin}.}\label{fig:setup}
\end{figure}

Moreover, there is {\em no central node} to aggregate the summaries computed by the individual nodes.
 The individual nodes exchange  information about the parameters of their measurement signals, rather than the raw signal itself, with their neighboring nodes  via a communication network. The communication network is specified by a directed graph $\G= (V, E)$, where $V$, $|V|=p$, denotes the set of nodes and $E$ denotes the set of directed edges.
 The neighbor set of the $g^{\rm th}$ node (sensor) is given by $\N_g = \{j: (g, j) \in E\}$.
We denote the local measurement available to node $g$  by $\Y_g$, where $\Y_g \subset \{\by_1, \by_2, \ldots, \by_q\}$ such that $\cup_{g=1}^p \Y_g = \{\by_1, \by_2, \ldots, \by_q\}$ and $\Y_g \cap \Y_j = \emptyset$ for $g,j \in \{1,2,\ldots,p\}$and $g \neq j$.  A schematic diagram of the setting is given in Figure~\ref{fig:decentral}. The goal is to recover the matrix $\bXs$ from the measurements of $p$ sensors in a fully decentralized and federated manner, specifically when $m <<n$. We refer to this problem as the  {\em Decentralized Low Rank Compressive Sensing (DLRCS) problem}.

A scalar representation of \eqref{eq:PR-1} is
\begin{equation}\label{eq:PR-2}
\by_{ik}:= \langle \ba_{ik}, ~\bxs_k \rangle, ~i=1,2,\ldots,m,~ k=1,2,\ldots, q.
\end{equation}
In Eq.~\eqref{eq:PR-2}, $\by_{ik}$ is the $i^{\rm th}$ entry of $\by_k \in \mathbb{R}^{m}$ and $\ba_{ik}^{\top} $ is the $i^{\rm th}$ row of the matrix $\bA_k$. We note that, the measurements are not global, since each measurement, $\by_{k}$, is a function of a particular column of $\bX^{\star}$, i.e., $\bxs_k$, rather the full matrix $\bX^{\star}$.
In other words, the measurements are global for each column  but not across the different columns. We thus need the following incoherence assumption to enable correct interpolation across the different columns \cite{lrpr_it}. This was introduced in \cite{matcomp_candes} for LR Matrix Completion (LRMC) which is another LR problem with non-global measurements, but its model is symmetric across rows and columns.

Let us denote the reduced (rank $r$) Singular Value Decomposition (SVD) of the rank-$r$ matrix $\bXs$ as
\begin{equation}\label{eq:SVD}
\bXs \svdeq  \bUs~ \Sigma^{\star}~ {\bV^{\star}}^{\top}.
\end{equation}
Here $\bUs \in \mathbb{R}^{n \times r}$ and $\bV^{\star} \in \mathbb{R}^{q \times r}$ are rank-$r$  orthonormal matrices, i.e., tall matrices with orthonormal columns. We use $\sigma_{\max}$  to denote the maximum  singular values of $\Sigma^{\star}$ and $\sigma_{\min}$ its minimum singular value. Thus $\kappa = \sigma_{\max}/\sigma_{\min}$ is the condition number of $\Sigma^\star$ (note that since $\bXs$ is rank deficient, its condition number is infinite).   We define the notations $\bB:=\bV^{\top}$ and $\tilde{\bB}:= \Sigma \bV^{\top}$. Thus $\bXs \svdeq  \bUs~ \Sigma^{\star}~ \bBs = \bUs~ \tilde{\bB}^{\star}$. In our approach, we will recover columns of $\tilde{\bB}^{\star}$, denoted as $\tilde{\bb}^{\star}_k$, individually.

\begin{assume}[Right singular vectors' incoherence]\label{assime:right_incoh}
We assume that  $\max_k \|{\bf b}^{\star}_k\| \le \mu \sqrt{r/q}$. Treating the condition number $\kappa$  of $\Sigma^{\star}$ as a constant, up to constants, this is equivalent to requiring that $\max_k \|\bxs_k\|^2 \le \tilde\mu \sum_{k=1}^q \|\bxs_k\|^2/q$ for a constant $\tilde\mu$ that can depend on $\kappa$.
\end{assume}

We  assume that the communication network $\G$ is strongly connected\footnote{A directed graph is said to be strongly connected if for each ordered pair of vertices $(v_i,v_j)$ there exists an elementary path from $v_i$ to $v_j$ \cite{Die:00}.} and symmetric, i.e., if node $g$ communicates
with node $j$, then node $j$ also communicates with node $g$. Consequently, the
in-degree equals out-degree for all nodes and the network is balanced \cite{olfati2004consensus}.   Hence  the following  assumption holds. 

\begin{assume}\label{assume:sc}
The  directed graph of the communication network $\G$ is strongly connected and balanced.
\end{assume}


\subsection{Notation}
We denote the Frobenius norm as $\norm{\cdot}_F$,  the induced $\ell_2$ norm as $\norm{\cdot}$, and the (conjugate) transpose of a matrix $\bZ$ as $\bZ^{\top}$. We use $\be_k$ to denote the $k^{\rm th}$ canonical basis vector and $h \in [d]$ for $h \in \{1,2, \ldots, d\}$ for some integer $d$. We define the Subspace Distance (SD) measure between two matrices $\bU_1$ and $\bU_2$ as $SD(\bU_1, \bU_2):= \norm{(I-\bU_1 \bU_1^{\top})\bU_2}_F$, where $I$ is the identity matrix. Note that, for two $r$-dimensional subspaces, $SD(\cdot, \cdot)$ is the $\ell_2$ norm of the sines of the $r$ principal angles between span$(\bU_1)$ and span$(\bU_2)$ and is a measure of distance between the two subspaces.

\section{Proposed Algorithm: DeF-GD }\label{sec:algos}
In this section, we  present the proposed fully decentralized federated algorithm for solving the  DLRCS problem.
We would like to find a matrix $\bX= [\bx_1, \bx_2, \ldots, \bx_q]$ that minimizes $f(\bX): = \sum_{k=1}^q ||\by_k - \bA_k {\bf x}_k||^2$ subject to the constraint that its rank is $r$ or less, in a fully decentralized and federated manner. The pseudocode for the proposed algorithm is given in Algorithm~\ref{alg:DeFe}. Algorithm~\ref{alg:DeFe} integrates a projected GD algorithm  with a consensus algorithm. The projected GD serves the matrix recovery part and the consensus algorithm serves the decentralized aggregation in a federated manner.
We first present the details for projected GD  and then present the details of consensus-based projected GD.

\subsection{Main idea of the centralized projected GD algorithm \cite{lrpr_gdmin}}
To recover matrix $\bX$, we write  $\bX = \bU \bB$ where $\bU$ is $n \times r$ and $\bB$ is $r \times q$ and do alternating projected GD on $\bU$ and $\bB$.  We use projected GD for updating $\bU$ (one GD step followed by projecting onto the space of orthonormal matrices); the projection step is needed to ensure the norm of $\bU$ does not keep increasing over iterations). For each new estimate of $\bU$, we solve for $\bB$ by minimizing over it keeping $\bU$ fixed. Because of the specific asymmetric nature of our measurement model, the $\min$ problem for columns of $\bB$ is decoupled. Thus the minimization over $\bB$  only involves solving $q$ $r$-dimensional Least Squares (LS) problems, in addition to also first computing the $q$ matrices, $\bA_k \bU$, for use in the LS step. Thus the time needed is only $O(q m r^2 + q mn r ) = O (mq nr)$. This is order-wise equal to the time needed to compute  gradient with respect to $\bU$, and thus, the per-iteration cost of GDmin is only $O(mqnr)$.


Notice that, for $m < n$, our problem is convex but not strongly convex. As a result GD starting from any arbitrary initialization may converge to {\em a} minimum but the minimum is not unique. Consequently, it is not guaranteed to converge to the true matrix that we want to recover. To address this issue, a class of approaches known as spectral initialization have been used frequently in the literature. The idea is to define a matrix that is close to a matrix whose top $r$ left or right singular vectors span the column span of the true $\bXs$ which is equivalent  to  a matrix  that is close to the top $r$ eigenvectors of $\bXs {\bXs}^{\top}$.
In our setting this involves computing  the matrix $\bU^{(0)}$ given in Algorithm \ref{alg:DeFeInit}.

\subsection{Decentralized and Federated Projected GD (DeF-GD)}
We note that the  measurements of the nodes are distributed as  $\Y_1, \ldots, \Y_p$, where $\Y_g \subset \{\by_1, \ldots, \by_q\}$, $\cup_{g=1}^p\Y_g = \{\by_1, \ldots, \by_q\}$ and $\Y_g \cap \Y_j = \emptyset$ for $j \neq g$. In a {\em decentralized federated} setting, the nodes  only share the parameters or estimates of the local updates, rather than the raw signal or the local measurements itself, with other nodes. Additionally, each node shares  the parameters or estimate of the local update   with the neighboring nodes only. As a result, a direct implementation of projected GD is not feasible. To propose a decentralized, federated version of the projected GD, we integrate projected GD with a consensus algorithm.

In each iteration of the algorithm, the nodes run a local projected GD  utilizing its local data. The parameters of the GD is then communicated with the neighboring nodes. Each node  aggregates its own local GD parameter with the neighbors' GD parameters and performs a distributed consensus until all nodes converge to the same GD parameter.  Once consensus is achieved, all nodes update their local estimate using the converged GD parameter in order to minimize the estimation error.  The projected GD iteration continues until all nodes converge to a global estimate with an acceptable  error tolerance. The convergence of the consensus algorithm is guaranteed when the communication graph is strongly connected and also the weight matrix is doubly stochastic and symmetric \cite{xiao2007distributed, olshevsky2009convergence}.

\begin{prop}[\cite{olshevsky2009convergence}]\label{prop:cons}
Let $\G$ be a strongly connected graph and suppose that each node of $\G$ performs a  distributed  linear protocol $z_g(t+1)= z_g(t)+ \sum_{j \in \N_g}{\bf W}_{gj}(z_j(t)-z_g(t))$. Then if the graph $\G$ is connected and ${\bf W}$ is doubly stochastic and symmetric, then  $\lim_{t \rightarrow \infty} z_g(t) = \dfrac{1}{p} \sum_{g =1}^p z_g(0)$ (average consensus), where $p$ is the number of nodes.
\end{prop}

In this work, we consider a strongly connected and balanced network $\G$,  and a doubly stochastic and symmetric  weight matrix $\bf{W}$. The consensus algorithm converges to the average value  by Proposition~\ref{prop:cons}. Below, we explain the details of our DeF-GD algorithm, including the initialization chosen for the projected GD.

\begin{algorithm}[t]
\caption{Pseudocode for distributed average consensus: \textsc{AvgConsensus} (${\bf D}_1, {\bf D}_2, \ldots, {\bf D}_p, {\bf W}, C$)\label{alg:consensus}}
\begin{algorithmic}
\State \textit {\bf Input:} Matrices ${\bf D}_1, \ldots, {\bf D}_p,$ where ${\bf D}_g \in \mathbb{R}^{u \times v}$ for $g \in \{1,\ldots,p\}$, Weight matrix ${\bf W} \in \mathbb{R}^{p \times p}$, iteration number $C$
\end{algorithmic}
\begin{algorithmic}[1]
\State Initialize ${\bf D}_g(0) \leftarrow {\bf D}_g$, for  $g=1,2,\ldots, p$
\For{$\tau=0$ to $C$}
\For {$g=1$ to $p$}
\State ${\bf D}_g (\tau+1) \leftarrow{\bf D}_g (\tau) +  \displaystyle\sum_{j \in \N_g} W_{gj}\Big( {\bf D}_j (\tau)- {\bf D}_g (\tau) \Big)$\label{step:avg}
\EndFor
\EndFor
\end{algorithmic}
\end{algorithm}

\begin{algorithm}
\caption{Pseudocode for federated initialization
\label{alg:DeFeInit}}
\begin{algorithmic}
\State \textit {\bf Input:} $\Y_g, \bA_k$, where $g=1,2,\ldots, p$ and $k=1,2,\ldots, q$
\end{algorithmic}
\begin{algorithmic}
\State \textit {\bf Output:} $\bU^{(0)}$, $\eta$
\end{algorithmic}
\begin{algorithmic}[1]
\State Initialize $\delta_g(0) \leftarrow \displaystyle\sum_{k \in [q]: \by_k \in \Y_g}\sum_{i=1}^m \by_{ik}^2$ for  $g=1,2,\ldots, p$ and $(\bU_g^{(0)})_0\leftarrow (\bU^{(0)})_0$, for  $g=1,2,\ldots, p$
\Return $\delta \leftarrow p\times$ {AvgConsensus} \Big({$\delta_1(0), \ldots, \delta_p(0), {\bf W}, C $}\Big)\label{step:cons-1}

\For{$\ell=0$ to $B$}
\For{$g=1$ to $p$}
\State $(\hat{\bU}_g^{(0)})_{\ell} \leftarrow  \displaystyle\sum_{k \in [q]: \by_k \in \Y_g}\dfrac{1}{m}\Big(\sum_{i=1}^m{\ba_{ik}\by_{ik}\mathbb{1}_{\{\by_{ik}^2 \leqslant 9 \delta / (mq) \}}} \Big)$ $\Big(\sum_{i=1}^m{\ba_{ik}\by_{ik}\mathbb{1}_{\{\by_{ik}^2 \leqslant  9 \delta / (mq) \}}} \Big)^{\top}(\bU_g^{(0)})_{\ell-1}$
\EndFor
\Return $\hat{\bU}^{(0)} \leftarrow p\times$ {AvgConsensus} \Big({$(\hat{\bU}_1^{(0)})_{0}, \ldots, (\hat{\bU}_p^{(0)})_{0}, {\bf W}, C $}\Big)\label{step:agg-2}

\State Obtain $\bU^{(0)}$ by QR, i.e., compute $\hat{\bU}^{(0)} \qreq \bU^{(0)} R^{(0)}$
\State  Set $(\bU_g^{(0)})_{\ell} \leftarrow \bU^{(0)}$, for all $g=1,2,\ldots, p$
\EndFor
\Return $\bU^{(0)}$
\Return $\eta \leftarrow 1/\lambda_{\max}( {R^{(0)}})$, $\lambda_{\max}(\cdot)$ denotes the largest eigenvalue
\end{algorithmic}
\end{algorithm}

We note that  the initialization step for Algorithm~\ref{alg:DeFe} also need to be done in a federated setting. As the measurements are distributed across the nodes and since the communication is federated, we will need a federated algorithm for initialization so that all nodes are initialized to a common value.  We explain this below and the pseudocode  of the initialization algorithm is presented in Algorithm~\ref{alg:DeFeInit}.

\subsubsection{Federated Initialization: Algorithm~\ref{alg:DeFeInit}}
 For federated initialization,  we use two  key steps, (i)~federated computation of the threshold of the indicator function and (ii)~federated Power Method (PM). To compute the threshold for the indicator function, each node, $g \in \{1, \ldots, p\}$, computes $\delta_g (0):= \displaystyle\sum_{k \in [q]: \by_k \in \Y_g} \sum_{i=1}^m \by_{ik}^2$, i.e., the squared sum of all the measurement available to node $g$. Each node then communicates this value with its neighboring nodes and performs a  distributed average consensus (step~\ref{step:cons-1} of Algorithm~\ref{alg:DeFeInit}). We present the subroutine code for distributed average consensus in Algorithm~\ref{alg:consensus}.

  \textsc{AvgConsensus} takes as input $u \times v$ matrices ${\bf D}_1, {\bf D}_2, \ldots, {\bf D}_p$ corresponding to $p$ nodes, a weight matrix ${\bf W} \in \mathbb{R}^{p \times p}$, where ${\bf W}$ is doubly stochastic and symmetric, i.e., ${\bf W}_{gj} = {\bf W}_{jg}$ and the maximum number of iterations  $C$. In each iteration, nodes update their values by taking weighted sum of its own and its neighbors' values (step~\ref{step:avg} of Algorithm~\ref{alg:consensus}).  Convergence of the \textsc{AvgConsensus} algorithm is guaranteed by Proposition~\ref{prop:cons}.

  In our algorithm we use consensus for scalar  values (i.e., $u=v=1$) and matrices. In the case of scalar, each node has a scalar value associated with it and the nodes communicate with neighbors to reach  consensus to the average value (e.g., step~\ref{step:cons-1} in Algorithm~\ref{alg:DeFe}). In the matrix case, each node is associated with a matrix and the nodes communicate with neighbors to reach consensus to the element-wise weighted average (e.g., step~\ref{step:agg-2} in Algorithm~\ref{alg:DeFe}), which is a direct extension of the scalar case. We note that, the number of iterations are chosen such that the values of the nodes are within an acceptable tolerance. We  obtain the threshold value of the indicator function using \textsc{AvgConsensus}. Once consensus of the  threshold value is achieved within a desired tolerance, a federated PM algorithm is executed using the converged threshold.

As explained in \cite{lrpr_gdmin}, we plan to initialize $\bU_0$   as the top $r$ left singular vectors of $\bX_0$, where
$$\bX_0:=\dfrac{1}{m}\Big(\sum_{k=1}^q \sum_{i=1}^m{\ba_{ik}\by_{ik}\mathbb{1}_{\{\by_{ik}^2 \leqslant 9 \delta / (mq) \}}} \Big),$$
where $\delta := \sum_{g=1} ^p\displaystyle\sum_{k \in [q]: \by_k \in \Y_g} \sum_{i=1}^m \by_{ik}^2 = p\times$  {AvgConsensus} ({$\delta_1(0), \ldots, \delta_p(0), {\bf W}, C $}).
This is equivalent to initializing  $\bU_0$   as the top $r$  eigenvectors of $\bX_0\bX_0^{\top}$. In order to compute the eigenvalues of $\bX_0\bX_0^{\top}$ in a federated and decentralized manner, we perform a federated Power Method (PM) \cite{lrpr_gdmin} followed by an average consensus.  In the federated PM, all nodes first jointly does a random  initialization $(\bU_g^{(0)})_0:= (\bU^{(0)})_0$ for all $g \in \{1,2,\ldots, p\}$, where $(\bU^{(0)})_0$ is a random matrix. Then, during each PM iteration, $\ell = 1,2, \ldots, B$, node $g$ computes
{\scalefont{0.85}{
\begin{equation*}\label{eq:init}
\displaystyle\sum_{k \in [q]: \by_k \in \Y_g}\hspace*{-1.5 mm}\dfrac{1}{m}\Big(\hspace*{-1.2 mm}\sum_{i=1}^m{\ba_{ik}\by_{ik}\mathbb{1}_{\{\by_{ik}^2 \leqslant 9 \delta/(mq) \}}} \Big)\dfrac{1}{m} \Big(\hspace*{-1.2 mm}\sum_{i=1}^m{\ba_{ik}\by_{ik}\mathbb{1}_{\{\by_{ik}^2 \leqslant 9 \delta/(mq) \}}} \Big)^{\top}\hspace*{-1.5 mm} (\bU_g^{(0)})_{\ell-1}
\end{equation*}
}}
 and communicates this information with its neighboring nodes.
Each node now aggregates its own and the neighbors' information using the \textsc{AvgConsensus} as in step~\ref{step:agg-2} of Algorithm~\ref{alg:DeFeInit}. The nodes perform a distributed consensus  until all nodes converge to the same $(\hat{\bU}_g^{(0)})_C$. Finally all nodes compute a QR factorization to obtain $\bU^{(0)}$ and proceeds to the next iteration of the federated PM. The outputs of Algorithm~\ref{alg:DeFeInit} are  ${\bU}_0$ and $\eta$ which serves as the initialization of the gradient decent and the step size for gradient decent, respectively, for the decentralized, federated LRCS algorithm, DeF-GD, presented in Algorithm~\ref{alg:DeFe}.

\subsubsection{Decentralized Projected Gradient Decent: Algorithm~\ref{alg:DeFe}}
Using the federated initialization, we propose a decentralized projected gradient descent algorithm to reconstruct the signal matrix $\bXs$.  Each node update its local estimation via a negative-gradient step, by combining the local gradient computed by the node using the  data available to the node, and the average of its neighbors' gradient estimates.

 Let $\bXs = \bUs \bBs$. We define the notations
\begin{eqnarray*}
f(\bU, \bB) &:=& \displaystyle\sum_{k=1}^q\displaystyle\sum_{i=1}^m(\by_{ik}-{\ba_{ik}^{\top}\bU\bb_k})^2 \mbox{~ and ~}\\
f_k(\bU, \bB) & :=& \displaystyle\sum_{i=1}^m(\by_{ik}-{\ba_{ik}^{\top}\bU\bb_k})^2.
\end{eqnarray*}

We initialize the $\bU$ matrix corresponding to the nodes, denoted as ${\bU_g}$, as $\bU^{(0)}$ computed in Algorithm~\ref{alg:DeFeInit}.  Then in each iteration of the DeF-GD algorithm (Algorithm~\ref{alg:DeFe}),  we update the gradient of node $g$, $\sum_{k: \by_k \in \Y_g} \nabla_{\bU_g}f_k(\cdot, \cdot)$, denoted as $\Psi_g$,  by one step of GD on $\bU_g$,  combined with a weighted average of the neighbors' information, i.e., $\Psi_j = \sum_{k: \by_k \in \Y_j} \nabla_{\bU_j}f_k(\cdot, \cdot)$ for $j \in \N_g$ (steps~\ref{step:gd} and~\ref{step:psi}). Once the local gradient updates of all nodes reach consensus  to a common $\Psi$ (step:\ref{step:cons-2}), the $\bU$ matrix is updated as $\hat{\bU}^+ = \bU - \eta \Psi$, where $\eta$ is the gradient step computed in Algorithm~\ref{alg:DeFeInit}. We then perform QR factorization to get a matrix with orthonormal columns. For each new $\bU$, we update $\bB$ by minimizing $f(\bU, \bB)$ over $\bB$. For a fixed $\bU$, we note that, the minimization of $f(\bU, \bB)$ over $\bB$ involves solving $q$ decoupled $r$-dimensional least squares problem.

\begin{algorithm}
\caption{Pseudo-code for proposed DeF-GD
\label{alg:DeFe}}
\begin{algorithmic}
\State \textit {\bf Input:} $\Y_g, \bA_k$, where $g=1,2,\ldots, p$ and $k=1,2,\ldots, q$
\State \textit {\bf Parameters:} GD step size $\eta$, number of iterations $T$, error tolerance $\gamma$
\end{algorithmic}
\begin{algorithmic}
\State \textit{\bf Output:} $\bU$
\end{algorithmic}
\begin{algorithmic}[1]
\State  Execute Algorithm~\ref{alg:DeFeInit} and obtain $\bU^{(0)}$
\State Initialize $t=1$, $\bU^{(0)}_g \leftarrow \bU^{(0)}$, for all  $g \in\{1,2,\ldots, p\}$\label{step:initialize}

\While{$t\leqslant T$ and Err $ > \gamma$}
\State $\bU \leftarrow \bU^{(t-1)}$
\For {$g=1$ to $p$}
\State Let $\bU_g \leftarrow \bU^{(t-1)}_{g}$
\For {$\by_k \in \Y_g$}
\State Set $(\bb_k)_t \leftarrow (\bA_k\bU_g)^{\dagger} \by_k$
\State Set $(\bx_k)_t \leftarrow \bU_g(\bb_k)_t$
\State Compute $\nabla_{U_g} f_k(\bU_g, (\bb_k)_t) = \displaystyle\sum_{i=1}^m (\by_{ik}-\ba_{ik}^{\top}\bU_g(\bb_k)_t)\ba_{ik}(\bb_k)_t^{\top}$ \label{step:gd}
\EndFor
\State ${\Psi}_g \leftarrow \displaystyle\sum\limits_{k\in [q]:\by_k \in \Y_g}\nabla_{U_g}f_k(\bU_g, (\bb_k)_t)$\label{step:psi}
\EndFor

\Return $\Psi \leftarrow$ {AvgConsensus} \Big({$\Psi_1, \Psi_2, \ldots,  \Psi_p, {\bf W}, C $}\Big)\label{step:cons-2}

\State Set $\hat{\bU}^{+} \leftarrow \bU - \eta \Psi$

\State Obtain $\bU^+$ by QR, i.e., compute $\hat{\bU}^+ = \bU^+ R^+$
\State Set $\bU^{(t)} \leftarrow \bU^+$
\State Err $\leftarrow \mbox{SD}(\bU^{(t)},\bU^{(t-1)} ) $

\EndWhile

\end{algorithmic}
\end{algorithm}

\subsection{Discussion}
We note that the  decentralized GD approach proposed  in \cite{nedic2009distributed} (and follow up works)  for standard GD is not applicable for DLRCS.  In DLRCS, we use GD to update estimates of the column span of the true matrix $\bXs =\bUs \bBs$, i.e., span of columns of $\bUs$. The $n \times r$ matrix $\bUs$ is unique only up to right multiplication by an $r \times r$ rotation matrix since $\bUs \bBs = \bUs {\bf Q}{\bf Q}^{-1} \bBs$, where ${\bf Q}$ is a rotation matrix.
Consequently, in Algorithm~\ref{alg:DeFe}, we use projected GD to update the subspace estimates $\bU$; run one step of GD  with respect to the cost function followed by projecting the output onto the set of matrices with orthonormal columns via QR decomposition.
The approach of \cite{nedic2009distributed}  designed for standard GD  cannot be used for updating $\bU$ because it involves averaging the partial estimates $\bU_g$,  $g \in \{1,2,\ldots, p \}$, obtained locally at the different nodes. However, since $\bU_g$'s are subspace basis matrices, their numerical average will not provide a valid ``subspace mean'' \footnote{To compute the subspace mean of $\bU_g$'s w.r.t. the subspace distance $SD(.,.)$, one would need to solve $\min_{\bar{\bU}}  \sum_g SD^2(\bU_g, \bar{\bU}) $. This cannot be done in closed form and will require an expensive iterative algorithm.}.
%

\section{Simulations}\label{sec:sim}
\begin{figure*}[h]
\begin{subfigure}[b]{0.49\textwidth}
\centering
\vspace{-1.2in}
\includegraphics[width = 1.1\textwidth, height =4in]{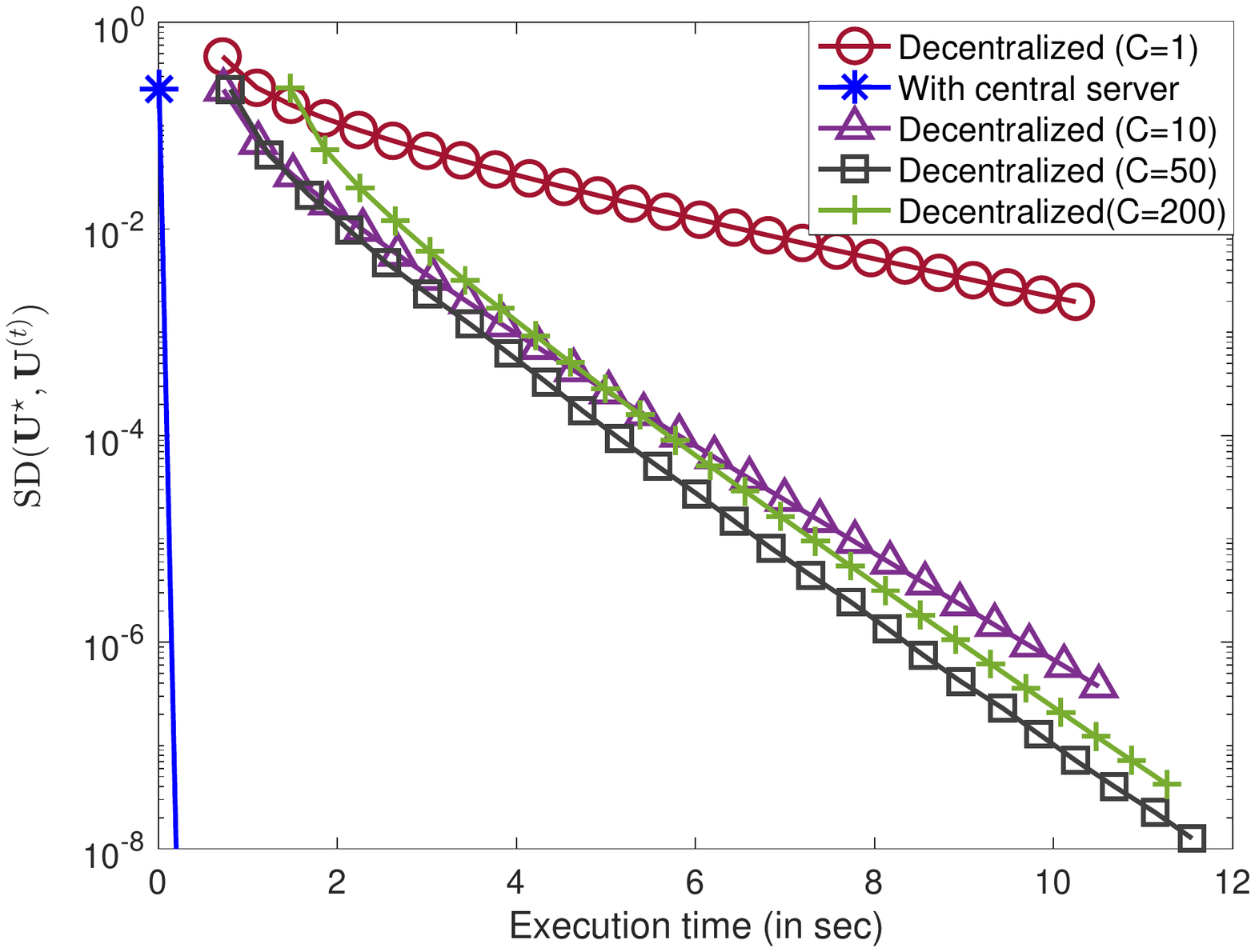}
\vspace{-1.2in}
\caption{}\label{fig:q4n}
\end{subfigure}
\begin{subfigure}[b]{0.49\textwidth}
\centering
\vspace{-1.2in}
\includegraphics[width = 1.25\textwidth, height =4in]{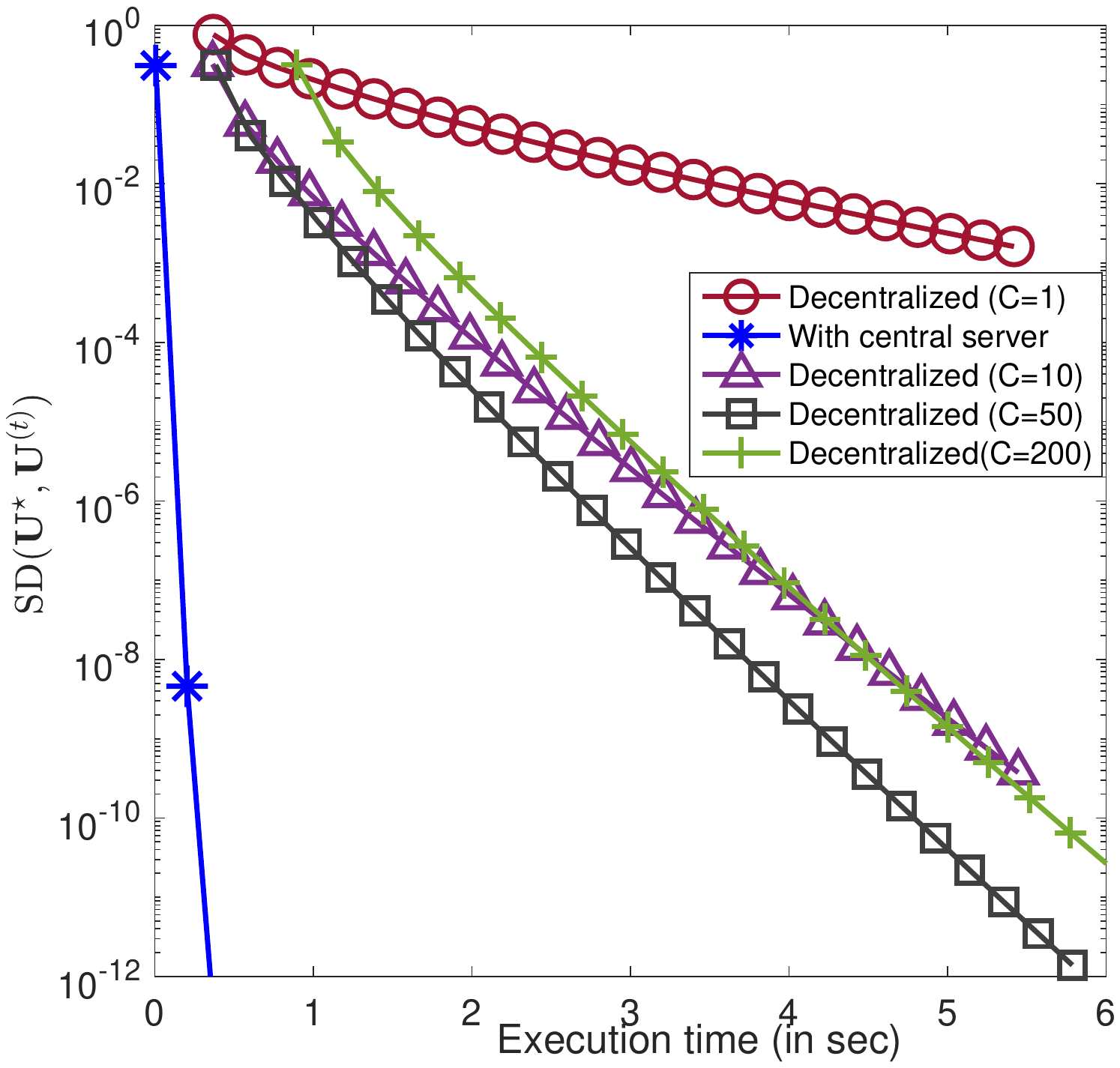}
\vspace{-1.2in}
\caption{}\label{fig:q2n}
\end{subfigure}
\caption{Error versus execution time plot with time in seconds. We  compare performance our fully decentralized algorithm (DeF-GD) by varying the number of iterations of  AvgConsensus as $C=1, 10, 50$, and $200$. We also compare DeF-GD   with the GDmin algorithm in \cite{lrpr_gdmin}, which is the memory efficient existing approach with guarantees  when there is a central server. In Figure~(\ref{fig:q4n}), $n=100$, $r=4$, $q=400$,  $m=40$, and $p=20$ and in Figure~(\ref{fig:q2n}), $n=100$, $r=4$, $q=200$,  $m=40$, and $p=20$.}\label{fig:Q}
\end{figure*}
In this section, we present the numerical validation of the proposed algorithm. We note that all the experiments were done using MATLAB.
 The communication network $\G$ and the dataset $\bA_k$'s and $\by_k$'s were generated randomly.

\newcommand{\prb}{\text{prob}}
We simulate the network as an Erd{\H o}s R\'{e}nyi  graph with $p$ vertices and with probability of an edge between any pair of nodes being  $\prb$. This means that there is an edge between any two nodes (vertices) $i$ and $j$ with probability $\prb$ independent of all other node pairs. For such a graph, if $\prb > (1+\zeta ) \log p/p$, then, for large values of $p$, with high probability (w.h.p.), the graph is strongly connected, i.e., Assumption~\ref{assume:sc}) holds. The probability that this holds goes to one as $p \rightarrow \infty$.
Also, if $\prb < (1+\zeta ) \log p/p$, then, for large values of $p$, w.h.p., the graph is not strongly connected. Since the guarantees are not deterministic, for a particular simulated graph, we used the {\em conncomp} function in MATLAB to verify that the graph is strongly connected.

We generated the data for our experiment as follows. We note that,  $\bXs = \bUs \tilde{\bB}^{\star}$ , where $\bUs$ is an $n \times r$ orthonormal matrix. We generate the entries of $\bUs$ by orthonormalizing an i.i.d standard Gaussian matrix. Similarly, the entries of  $\tilde{\bB}^{\star} \in \mathbb{R}^{r \times q}$ are generated from a different  i.i.d Gaussian distribution. The matrices $\bA_k$s
 were i.i.d. standard Gaussian. We performed three experiments on the generated dataset. (1)~Variation of the estimation error, denoted as $SD(\bUs, \bU^{(t)})$, where $\bUs$ is the actual matrix and $\bU^{(t)}$ is the estimate returned by our algorithm at iteration $t$, with respect to the time taken for execution for different values of $q$.   (2)~Variation of the estimation error, $\norm{\bXs-\bX^{(t)}}_F/\norm{\bXs}_F$, where $\bXs$ is the actual data matrix and $\bX^{(t)}$ is the estimate of $\bXs$ corresponding to the output of the algorithm at iteration $t$, with respect to the time taken for execution for different values of edge probability in the network $\G$.  (3)~Variation of the estimation error, $\norm{\bXs-\bX^{(t)}}_F/\norm{\bXs}_F$,  with respect to the time taken for execution for different values of edge probability and $C$. All the experiments were performed on  100 independent trials, i.e., each point in the experiment plot was averaged over 100 different samples of $\bA_k$s.

\noindent{\em \bf Experiment~1:} For this experiment, we generated  the communication network $\G$  such that there exists a link between two nodes with probability  $0.5$. Thus the communication network is an Erd{\H o}s R\'{e}nyi graph with edge probability 0.5. We we plot the matrix estimation error (at the end of the iteration) $SD(\bUs, \bU^{(t)})$ and the execution time-taken (until the end of that iteration) on the y-axis and x-axis, respectively. The parameters chosen for this experiment are: $n=100$, $r=4$,  $m=40$, and $p=20$. We provide results of the DeF-GD algorithm for four different values of the consensus iteration; (i)~$C=1$, (ii)~$C=10$,   (iii)~$C=50$, and  (iv)~$C=200$, when $q=400$ and $q=200$. We ran the DeF-GD algorithm for these cases and also implemented the GDmin algorithm given in \cite{lrpr_gdmin} for the centralized case where there is a central server that performs all the aggregation. In the GDmin algorithm all nodes send their gradients to the central server, and the central server aggregates the data.

The experimental results are presented in Figure~\ref{fig:Q}, where Figures~\ref{fig:q4n} and~\ref{fig:q2n} correspond to $q=400$ and $q=200$, respectively. From the experiments, we notice that, for a certain range of  $C$,  the rate of decay of  error increases as $C$ increases (for cases where $C=1, 10, 50$). However,  increasing $C$ beyond a certain value is not useful as it does not improve consensus further  and on the other hand, introduces additional computational overhead resulting in the increase of the execution time, as demonstrated in the case of $C=200$. We thus infer that,  the structure of the network and the partition of the data across the nodes, play a crucial role in deciding the number of iterations required for achieving consensus  and consequently  the  amount of computations required  to recover the data matrix.
We also notice that the error decays at a  faster rate for the GDmin algorithm, which is expected as the central server can directly aggregate parameters of all nodes. In the decentralized case, the nodes rely on the communication graph to exchange information and hence the error decays at a lower rate when compared to the the case with central server.

\begin{figure*}[h]
\begin{subfigure}[b]{0.49\textwidth}
\centering
\vspace{-0.9in}
\includegraphics[width = 1.05\textwidth, height =4in]{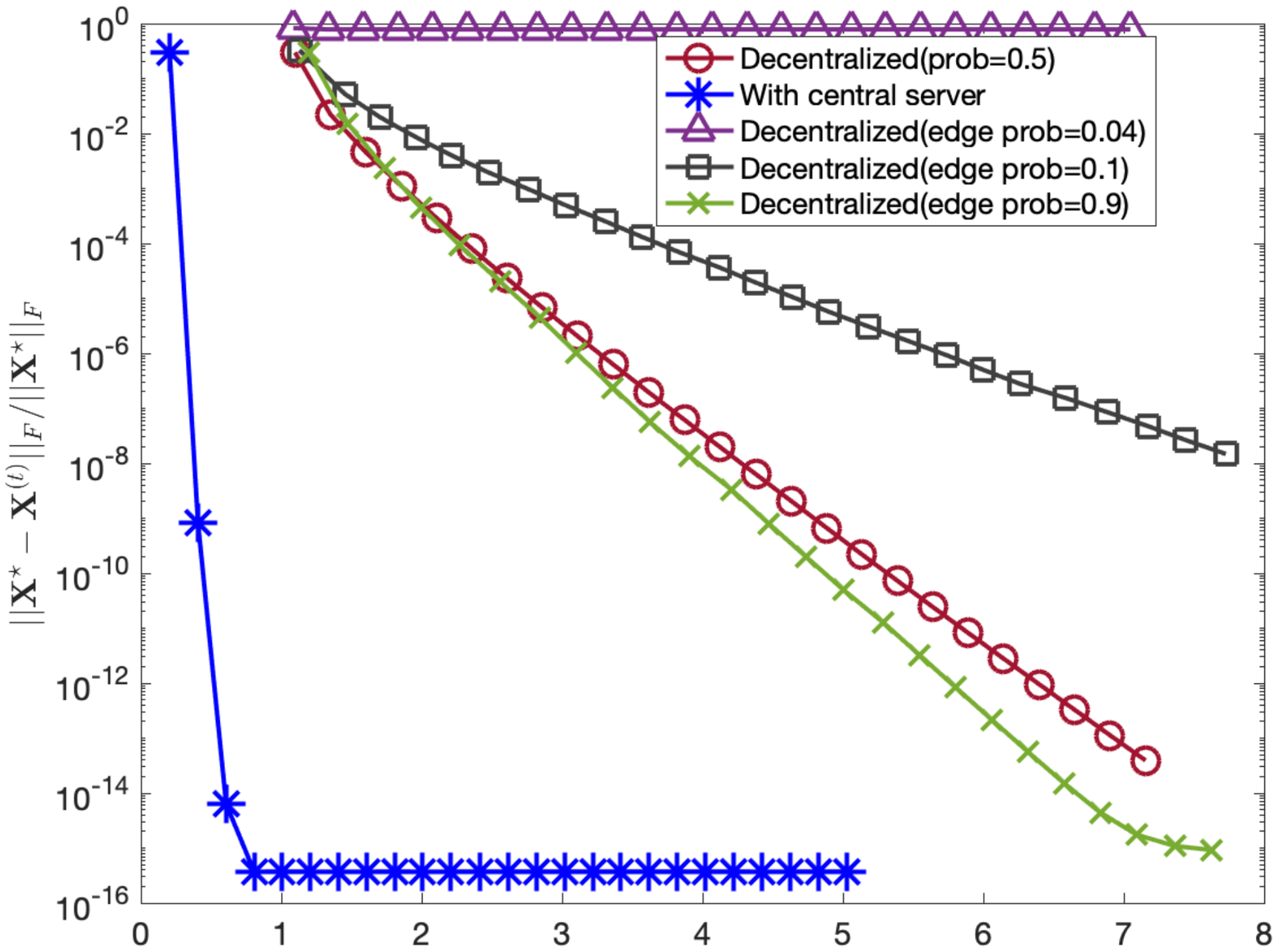}
\vspace{-1.1in}
\caption{}\label{fig:prob1}
\end{subfigure}
\begin{subfigure}[b]{0.49\textwidth}
\centering
\vspace{-1.2in}
\includegraphics[width = 1.15\textwidth, height =4.2in]{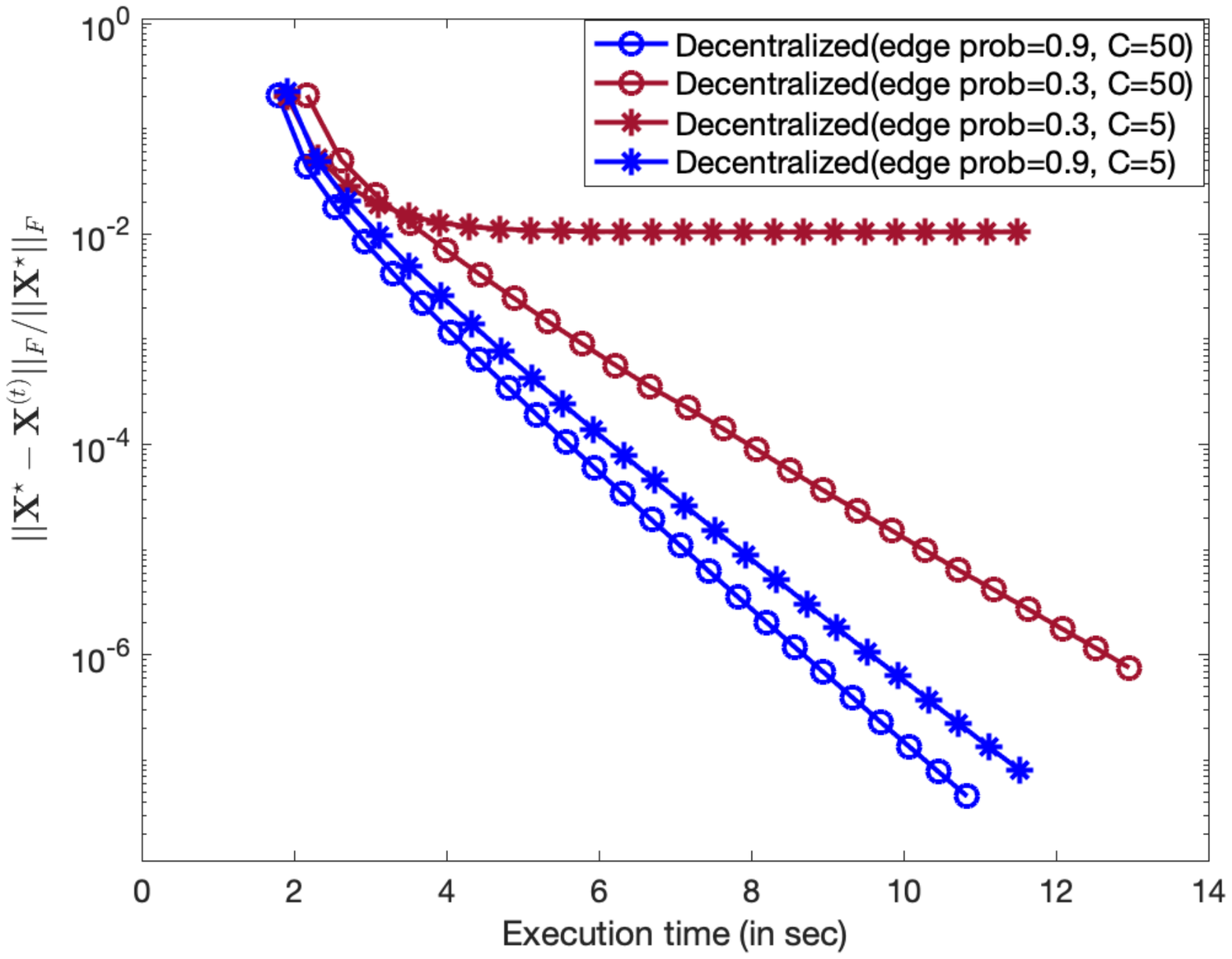}
\vspace{-1.2in}
\caption{}\label{fig:prob2}
\end{subfigure}
\caption{Error versus execution time plot with  time in seconds. In Figure~\ref{fig:prob1}, we  compare performance our fully decentralized algorithm (DeF-GD) by varying the probability of edge in the communication network $\G$ as $0.04, 0.1, 0.5$, and $0.9$. We also compare DeF-GD   with the GDmin algorithm in  \cite{lrpr_gdmin}, which is the memory efficient  approach with guarantees when there is a central server. The parameters used are:  $n=100$, $r=4$, $q=400$,  $m=40$,  $p=20$, and $C=50$. In Figure~\ref{fig:prob1}, we  compare the performance of the DeF-GD algorithm as the number of iterations of AvgConsensus and the connectivity of the network varies. The parameters used are:  $n=100$, $r=4$, $q=400$,  $m=40$,  and $p=20$.}\label{fig:prob}
\end{figure*}
\noindent{\em \bf Experiment~2:} For this experiment, we varied the edge probability of the   communication network $\G$ and analyze the estimation error. The parameters chosen for this experiment are: $n=100$, $r=4$,  $q=400$, $m=40$,  $p=20$, and $C=200$.  We  plot the matrix estimation error (at the end of the iteration) $\norm{\bXs-\bX^{(t)}}_F/\norm{\bXs}_F$ and the execution time-taken (until the end of that iteration) on the y-axis and x-axis, respectively. We provide results of the DeF-GD algorithm for four different values of the edge probability; (i)~$0.04$, (ii)~$0.1$,   (iii)~$0.5$, and (iv)~$0.9$. We note that, for edge probability $0.04$, the network $\G$ was not connected and this explains the reason for no decay in the error. On the other hand, for edge probability values $0.1, 0.5$, and $0.9$, the resulting network was connected.
We compared the  performance of the DeF-GD algorithm   with the  GDmin algorithm given in \cite{lrpr_gdmin}, where there is a central server. The experimental results are presented in Figure~\ref{fig:prob1}. From the results, as expected, our decentralized algorithm gives lower error and faster convergence when the network is a well connected graph. From  Figure~\ref{fig:prob1} we  infer that the decay rate of error increases as the probability of  an edge in the network increases. As a result, the convergence rate of the DeF-GD algorithm improves, and the gap between the centralized approach and the decentralized approach decreases as the connectivity of the  network increases.

\noindent{\em \bf Experiment~3:} The aim of this experiment was to analyze the behavior of the algorithm with respect to consensus iterations $C$ and edge probability. Here we vary both the edge probability of the   communication network $\G$ and the consensus iteration count $C$, and then analyze the estimation error. The parameters chosen for this experiment are: $n=100$, $r=4$,  $q=400$, $m=40$, and  $p=20$.  We  plot the matrix estimation error (at the end of the iteration) $\norm{\bXs-\bX^{(t)}}_F/\norm{\bXs}_F$ and the execution time-taken (until the end of that iteration) on the y-axis and x-axis, respectively. We provide results of the DeF-GD algorithm for four different cases; (i)~edge probability=$0.3$ and $C=5$, (ii)~edge probability=$0.3$ and $C=50$,   (iii)~edge probability=$0.9$ and $C=5$, and (iv)~edge probability=$0.9$ and $C=50$. We note that, for all the cases, the resulting network is connected. From Figure~\ref{fig:prob2}, for lower value of edge probability i.e., $0.3$, the error decay rate changes considerably with change in the number of consensus iteration $C$. However for the case when edge probability is $0.9$ (i.e., well connected graph), the effect of the number of consensus iteration on the decay rate of the error is small. We thus infer that the trade-off between the consensus accuracy and computational overhead varies depending on the connectivity structure of the network. While there is a considerable improvement in accuracy with consensus computations in the case of networks with fewer edges, the improvement in accuracy with consensus computations i decreases as the connectivity of the network increases.

\section{Conclusion}\label{sec:con}
In this paper we studied  the Low Rank Compressive Sensing (LRCS) problem  in a fully decentralized  setting, where the measurement signals are distributed across a set of nodes that are allowed to exchange their information only with a pre-specified set of neighboring nodes. We referred to this  problem as the Decentralized Low Rank Phase Retrieval (DLRCS) problem. We considered the federated setting of the DLRCS problem where the nodes only share the parameters of their local estimate rather than the raw signal itself. For solving the DLRCS problem, we proposed a fully decentralized, federated algorithm, referred to as DeF-GD. Our algorithm incorporated a projected gradient decent to serve the matrix recovery part and an average consensus algorithm for achieving the collaboration of nodes. We validated the effectiveness of our algorithm on randomly generated synthetic data and  compared with the existing memory efficient  approach in \cite{lrpr_gdmin}, which addresses the case when there is a central server in the system. We plan to investigate convergence guarantees of the proposed DeF-GD algorithm as part of our future work.

 \bibliographystyle{myIEEEtran}
 \bibliography{tipnewpfmt_kfcsfullpap.bib}

\end{document}